%% file: main.tex
\documentclass[twocolumn, conference, 10pt]{IEEEtran}

\usepackage{graphicx, pifont}
\usepackage{subfigure}
\usepackage{mathtools}
\usepackage{tikz}
\usepackage{soul}
\usepackage{cite}
\usepackage{changes}
\usepackage{physics}
\usepackage{amsmath,amssymb}
\usepackage{algorithm}
\usepackage{algpseudocode}
\usepackage[toc,page]{appendix}
\usepackage{wrapfig}
\usepackage[makeroom]{cancel}
\usepackage{tablefootnote}
\usepackage{comment}
\usepackage[english]{babel}
\usepackage[utf8]{inputenc}
\usepackage{boxedminipage}
\usepackage{textcomp}
\usepackage{xcolor, wrapfig}
\usepackage{fancyhdr}
\usepackage{tabularx}
\usepackage{makecell}
\presetkeys{todonotes}{disable}{}

\IEEEoverridecommandlockouts
\usepackage{listings}

\definecolor{mygreen}{RGB}{28,172,0}
\DeclareMathOperator\arctanh{arctanh}

\algnewcommand\algorithmicforeach{\textbf{for each}}
\algdef{S}[FOR]{ForEach}[1]{\algorithmicforeach\ #1\ \algorithmicdo}
\let\oldtodo\todo
\renewcommand{\todo}[1]{\oldtodo[inline, color=yellow]{{\bf TODO:} #1}}
\newcommand{\fixme}[1]{\fcolorbox{black}{red}{\color{yellow}{\bf FIXME:} #1}}
\newcommand{\note}[1]{\oldtodo[inline, color=green]{{\bf Note:} #1}}
\newcommand{\donotshow}[1]{}

\def\eg{{\em e.g.}, }

\newcommand{\red}[1]{{\color{red}#1}}
\newcommand{\blue}[1]{{\color{blue}#1}}

\newcommand*\xor{\oplus}

\graphicspath{{FIGS/}}

\title{Efficient LDPC Decoding using Physical Computation
}
\author{
\IEEEauthorblockN{Uday Kumar Reddy Vengalam, Andrew Hahn, Yongchao Liu, Anshujit Sharma, Hui Wu, and Michael Huang}
\IEEEauthorblockA{Department of Electrical and Computer Engineering, 
University of Rochester \\
\{uvengala, ahahn7, yongchao.liu, ashar36, hui.wu,
michael.huang\}@rochester.edu}
}

\begin{document}
\maketitle
\pagestyle{plain}

\input{Sections/0abstract}
\input{Sections/1intro.tex}

\input{Sections/2background.tex}

\input{Sections/3proposed.tex}

\input{Sections/4evaluation.tex}

\input{Sections/5conclusion.tex}

\small
\bibliographystyle{IEEEtran}
\bibliography{main}

%\newpage
%\input{LDPC codes/Sections/7notes}
\end{document}

%% file: Sections/0abstract.tex
\begin{abstract}\label{sec:abstract}

Due to 5G deployment, there is significant interest in LDPC decoding.
While much research is devoted on efficient hardwiring of algorithms 
based on Belief Propagation (BP), it has been shown that LDPC decoding can be formulated as a
combinatorial optimization problem, which could benefit from significant
acceleration of physical computation mechanisms such as
Ising machines. This approach has so far resulted in poor
performance. This paper shows that the reason is not fundamental but
suboptimal hardware and formulation. A co-designed Ising machine-based 
system can improve speed by 3 orders of magnitude. As a result,
a physical computation approach can outperform hardwiring
state-of-the-art algorithms. In this paper, we show such an
augmented Ising machine that is 4.4$\times$ more energy efficient than
the state of the art in the literature.

\begin{comment}
\begin{itemize}
    \item LDPC codes are important ECC codes for its ability to reach the Shannon limit
    \item they have been standardized for many different applications
    \item The decoding of LDPC codes is a combinatorial optimization problem
    \item Ising machines have shown to solve COPs very effectively
    \item major problem: to map cop with constraints on ising machines needs to be converted into a QUBO which requiers some extra spins
    \item this makes the state space for search increase exponentially
    \item with changes to the hardware of BRIM we can directly map decoding LDPC without conversion
    \item this leads to design of LDPC decoder with \blue{118.6 Gbits/sec} through, \blue{158.24mW} power consumption.
    \item this design helps in reduction of energy consumption per bit by an order of magnitude compared to other state of art LDPC decoders
    
\end{itemize}
\end{comment}

\end{abstract}


%% file: Sections/1intro.tex
\section{Introduction}

Error-correcting codes (ECCs) are essential in detecting and correcting
errors in modern communication and storage infrastructures. The central idea
of any ECC is to encode redundant information into the message sent. 
The redundancy allows the receiver to deal with a limited number
of errors induced by the communication channel or storage media. There
have been many ECCs introduced in literature, such as Hamming codes~\cite{hamming1950error}, convolutional codes~\cite{viterbi1971convolutional},
polar codes~\cite{arikan2008performance}, turbo codes~\cite{berrou1993near},
etc. However, among them, \emph{low density parity check} (LDPC)
codes~\cite{gallager1962low} have become more popular due to their
ability to achieve the Shannon limit~\cite{mackay1999good}. Recently
they have been introduced in the 5G-NR standard along with Polar
codes~\cite{bae2019overview}.

%\todo{ Ideally we will also summarize how many chips/designs along the line of OMS have been published with a range of performance of these chips.}

%\todo{I vaguely recall there's some notion that LDPC decode costs half of all energy in the entire chain. Do you remember that? Either way please go dig out a citation.}
Decoding LDPC is at the moment computationally intensive. 
Perhaps the most popular approach to LDPC decoding 
is a Belief Propagation (BP) type algorithm. More specifically, more than 200 ASIC
designs have been discussed in literature~\cite{shao2019survey,soton399259} 
all using some variant of
the Min-Sum algorithm~\cite{fossorier1999reduced}, where an
approximation is applied to the canonical BP algorithm to reduce 
computational intensity. 
An alternative approach to LDPC decoding is to treat it as a 
combinatorial optimization problem (COP). 
Algorithmically, Min-Sum variants are far more efficient 
than a typical COP solver when targeting
real-world LDPC decoding. However, the latter can benefit from 
orders-of-magnitude acceleration from emerging hardware
Ising machines~\cite{afoakwa.hpca21, inagaki2016cim, DBLP:journals/corr/abs-1903-07163}. 
Indeed, one such design using D-Wave's quantum annealer was found
to outperform an FPGA-based implementation of a variant of BP~\cite{kasi2020towards}.

\begin{comment}
\red{A natural question is whether the latter approach (formulating the 
problem as a COP and leverage a hardware annealer) 
is a more promising approach. The answer is a qualified yes.
While the rest of the paper discusses the many details, the main takeaways are:
\begin{itemize}
    \item For a typical LDPC decoding problem, Min-Sum style algorithm is 
    far more efficient (by about \fixme{1000x}) than solving it as a blackbox COP 
    for von Neumann architectures.
    \item Hardwiring Min-Sum in an ASIC can achieve \fixme{$10^3$} improvement in speedups. Similarly, a state-of-the-art hardware annealer can 
\end{itemize}

word under some model of noise behavior 

a generic optimization approach of using an annealer.

LDPC decoding can also be formulated as 
a constrained combinatorial optimization problem. Consequently, special-purpose annealers can be employed to perform decoding.}
\end{comment}

Nevertheless, LDPC decoding using Ising machines is not yet competitive with the state-of-the-art
ASIC designs hard-wiring Min-Sum algorithm.
Depending on the details of the custom hardware, these ASICs can expect a decoder throughput of 
roughly 0.13 - 271Gbps, with a power consumption of 5mW - 12W.
Using D-Wave hardware to solve an LDPC decoding problem today can only
achieve 21 Mb/s throughput, not to mention the kilo-watt power consumption.
Of course, part of that is specific to D-Wave hardware. But there are intrinsic
reasons as well. In this paper, we change this picture and show 
an efficient LDPC decoder based on physical computation
with the following contributions: \begin{enumerate}
\item \textbf{Fundamental analysis:} we show an important factor prohibiting out-performance is in 
current problem formulation when mapping to an Ising machine. 

\item \textbf{Co-designed architecture:} We address this problem by 
proposing a novel, co-designed Ising machine architecture that allows better expressivity of the target problem.

\item \textbf{Evaluation:} Using detailed simulation, we show 
decoding throughput of 81 Gbits/sec and a power consumption of 158.24mW, representing at least 4 times reduction in energy compared to state-of-the-art decoder ASICs for 5G standards.

\end{enumerate}

%% file: Sections/2background.tex
\section{Background and Related Work}
\label{sec:background}

We first discuss the communication model assumed. We then discuss LDPC codes  
and the Ising model.

\subsection{Communication model}
In a general communication system,  
a message $M$ is first encoded using an encoder into a code word $C$. This code 
word is modulated using a modulator into the signal to be sent. While being
transmitted, the signal gets affected by noise present in the channel.
The received signal is then demodulated and decoded to recover the original
message $M$, hopefully error free.

\begin{figure}[htp]
    \centerline{\includegraphics[width=0.50\textwidth]{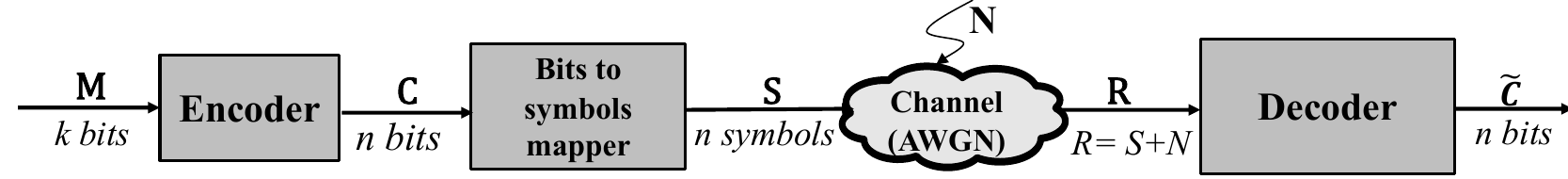}}
    \caption{Schematic diagram of communication system.}
    \label{fig:communication_digaram}
\end{figure}

%In this paper, we assumed a discrete communication system (Figure~\ref{fig:communication_digaram}),  where a $k$-bit binary message $M$ is  encoded into an $n$-bit code word ($C$) using a ($k$, $n$) LDPC code. This code word is modulated using the \emph{binary phase shift keying} (BPSK) modulation scheme, transmitted over a channel with additive white Gaussian noise (AWGN), and demodulated into received signal where $R_i=(1-2C_i)+\mathcal{N}(0, 1/SNR)$. 
%\todo{Professor please take a look at this paragraph }

In this paper, we assumed a discrete communication system (Fig.~\ref{fig:communication_digaram}), 
where a binary message $M$ with $k$-bit is encoded into an $n$-bit code word $C$ using a ($k$, $n$) LDPC  code. This code word is converted into an $n$-bit vector ($S_i = 1-2C_i$) of symbols representing a \emph{binary phase shift keying} (BPSK) constellation. This $n$-bit vector of symbols is transmitted over a channel with additive white Gaussian noise (AWGN). At the receiver end, we would receive a signal where $R_i=S_i+N(1/SNR)$, $N$ represents the channel's noise.

\subsection{Low Density Parity Check Codes}
LDPC codes are a class of linear block codes. The parity bits (redundant
information) are obtained using bitwise XOR operation on a selected number of
message bits. These codes are specified by a generator matrix $G_{k \times
n}$ or a parity check matrix $H_{m \times n}$, each of which can be obtained
from the other using the following relation $H\xor G^\top =0$. Here $m$ is
the number of parity bits, $n$ is the number of bits of the code word, and
$k$ is the number of bits in the message ($k=n-m$). The ratio of the length of
the message ($k$) to the size of the code word ($n$) is called the \emph{rate}: 
$r = k/n$. For example, the 5G standard uses 1/3 and 1/5 rate codes. 
The generator matrix is used for encoding a message ($M$) into a code 
word ($C$) using the following equation $C = G^\top\xor M$.
At the receiver side the decoded message is obtained from both the received
message ($R$) and the parity check matrix. The name ``low density" comes from the
sparse structure of the parity check matrix. 

\begin{figure}[htbp]
    \centering
    \includegraphics[width=0.38\textwidth]{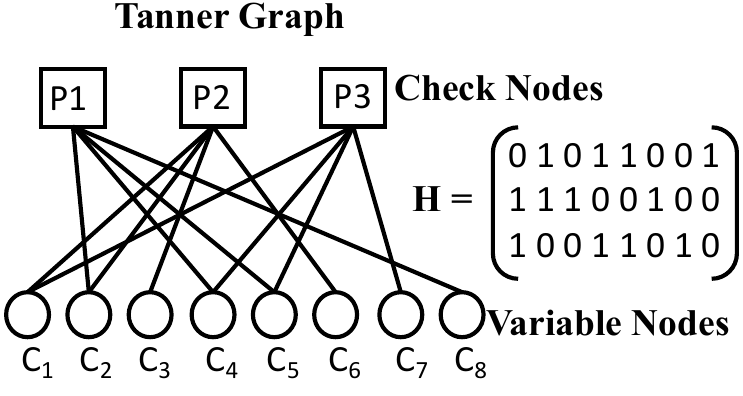}
    \caption{Tanner graph with corresponding parity check matrix $H$.}
    \label{fig:tg}
\end{figure} 

LDPC codes can be represented graphically using Tanner graphs. These graphs are bipartite, meaning that the graph nodes are separated into two distinctive sets, and edges only connect nodes of two different types. The two types of nodes in a Tanner graph are variable and check nodes. Fig.~\ref{fig:tg} is an example of a Tanner graph that is constructed using parity check matrix $H$ shown above. It consists of $m$ check nodes (the number of parity bits) and $n$ variable nodes (the number of bits in an encoded message). Check node $P_i$ is connected to variable node $C_j$ if the element $h_{ij}$ of $H$ is a 1. These graphical structures are handy for analysis and decoding purposes.

\subsection{Ising Model}\label{sec:ising_mode}

The Ising model is used to describe the energy of a system of coupled
spins. The spins have one degree of freedom and take one of two values ($+1$,
$-1$). The energy of the system is a function of pair-wise coupling of the
spins ($J_{ij}$) and the interaction ($h_i$)
of some external field ($\mu$) with each
spin. The resulting energy is as follows:

\begin{equation}
\label{eqn:Ising_w_field}
% H = -\sum_{(i<j)} J_{ij}\sigma_i\sigma_j - \mu \sum_i h_i\sigma_i
E = -\sum_{\mathclap{(i<j)}} J_{ij}\sigma_i\sigma_j - \mu \sum_{\mathclap{i}} h_i\sigma_i
\end{equation}

A physical system with its energy defined by the formula naturally tends towards low-energy
states. It is as if nature tries to solve an optimization problem with
Eq.~\ref{eqn:Ising_w_field} as the objective function, which is not a trivial
task. Indeed, the cardinality of the state space grows exponentially with the
number of spins, and the optimization problem is NP-complete: it is easily
convertible to and from a generalized max-cut problem, which is part of the
original list of NP-complete problems~\cite{Karp1972}.

Thus if a physical system of spins somehow offers programmable coupling
parameters ($J_{ij}$ and $\mu h_i$ in Eq.~\ref{eqn:Ising_w_field}), they can
be used as a special purpose computer to solve\footnote{Throughout
the paper, by ``solving" an optimization problem we mean the
searching for a good solution rather than necessarily finding the global optimum, 
equivalent to reaching the ground state.} optimization problems that
can be expressed in the form of an Ising formula (Eq.~\ref{eqn:Ising_w_field}). In fact, all
problems in the Karp NP-complete set have their Ising formula
derived~\cite{lucas2014ising}. Also, if a problem already has a QUBO (quadratic
unconstrained binary optimization) formulation, mapping to Ising formula is as
easy as substituting bits for spins, \eg $\sigma_i = 2b_i-1$.

Because of the broad class of problems that can map to the Ising formula,
building nature-based computing systems that solve these problems has
attracted significant attention~\cite{kim2010quantum,berloff2017realizing,
barends2016digitized,yamaoka201520k,bunyk2014architectural,king2018observation,
bohm2019poor,hamerly2019towards,pierangeli2019large}.

%% file: Sections/3proposed.tex
\section {Co-designed Ising Machine Architecture}
\label{sec:arch}

%In this section, we discuss about the algorithm change to QUBO formulation of decoding of LDPC codes and it's formulation.
\subsection{Fundamentals of QUBO formulation of LDPC decoding}
\label{sec:QUBO_formulation}

The goal of LDPC decoding is to reproduce the original message $M$ with low bit error rates. 
This is modeled as a combinatorial optimization problem with two objectives: \begin{enumerate}
    \item The decoded output $\tilde{C}$ is a code word (satisfies parity check).
    \item It should be close to the received message $R$.
\end{enumerate}
If we use the BPSK modulation scheme, the objective function can be represented by Eq.~\ref{eq::obj}
where $\alpha$ adjusts the relative emphasis of the two objectives above.\footnote{Choosing $\alpha > \sum_{i=1}^{n} 4|R_i|$ is suggested~\cite{tawada2020new} to \emph{guarantee} the decode output is always a code word. In our observations, this choice results in poorer results than using empirically selected values.} \begin{equation}
\label{eq::obj}
\begin{split}
    F =  \sum_{i=0}^{n-1}(R_i - (1-2\tilde{C_i}))^2 + \alpha (H \xor \tilde{C})
\end{split}
\end{equation}

%\todo{shouldn't there be a factor before $H \xor \tilde{C}$ to determine how important satisfying parity check.}
\begin{comment}
\begin{equation}
\label{eq::obj}
\begin{split}
    F = \sum_{i=0}^{n-1}(R_i - (1-2\tilde{C_i}))^2 \\
   with\ constraint\ that\  H \xor \tilde{C} = \hat{0}
\end{split}
\end{equation}
\end{comment}

\subsubsection{Conversion to QUBO formulation}
Fundamentally, because of the presence of the XOR operation, Eq.~\ref{eq::obj}
is not a quadratic formula (\eg QUBO). Therefore it can not be directly mapped
to an Ising machine. There are multiple ways of converting the formulation into
a quadratic form, but here we only discuss one, which uses minimum auxiliary
spins~\cite{tawada2020new}. The general idea is straightforward: Satisfying
parity check means that $H\xor \tilde{C}$ is an all-zero vector. This in turns
means that the inner product of each row of $H$ with $\tilde{C}$ is an
even number. This can be represented as $H_j \cdot \tilde{C} = 2L_j$, where
$H_j$ is the $j$-th row of the parity check matrix ($H$), and $L_j$ is any integer.
This condition can be represented using the following objective function:
\begin{equation}
    F_s = \sum_{j=0}^{m-1}(\sum_{i=0}^{n-1} H_{j,i}\tilde{C}_i - 2L_j)^2.
\end{equation}
Since the integer $L_j$ is a free variable, it needs to be
to be represented by auxiliary variables $y_{j,k}\in \{0,1\}$. 
Given a fixed LDPC code, $L_j$ is
bounded: $0 \leq L_j \leq \lfloor \frac{|H_j|}{2} \rfloor $, where $|H_j|$
is the number of 1’s in $H_j$. 
This can be done in a number of ways with different tradeoffs. One approach is
\emph{unary} (or one-hot) encoding, another is \emph{binary} encoding as follows.
\begin{equation}\label{eq:unary_binary}
\begin{split}
    \text{Unary encoding:\;}&  L_j = \sum_{k=0}^{\lfloor \frac{|H_j|}{2} \rfloor} y_{j,k} \\
    \text{Binary encoding:\;}&
             L_j = \sum_{k=0}^{\lfloor \log_2 (\frac{|H_j|}{2})\rfloor} 2^k y_{j,k}
\end{split}
\end{equation}
Thus, the objective function can be written as: 
\begin{equation}
\label{eq::obj_ising}
    F = \sum_{i=0}^{n-1}\left(R_i - (1-2\tilde{C_i})\right)^2 + \alpha F_s,
\end{equation}
and after collecting terms,
$F$ can be written in the following format: 
\begin{equation}
F = x^\top Qx +c 
\end{equation}
where $x$ represents all binary variables including those of $\tilde{C}$ and the auxiliary variables $y_{j,k}$. The matrix $Q$ is mapped onto Ising machines to find $x$ as a solution for the decoding problem.

This conversion is necessary for a standard Ising machine. But it requires
extra spins. In fact, binary encoding (Eq.~\ref{eq:unary_binary}) almost
doubles the total number of spins required and unary encoding requires even
more. Keep in mind that the size of an Ising model's state space scales with
$2^n$ where $n$ is the number of spins. The added auxiliary spins not only
increase the hardware resource demand, but also has significant implications
on an Ising machine's ability to navigate through the solution space and thus
on the solution quality (see Sec.~\ref{sec:solution_quality}). To address the 
issue, we propose a new co-designed Ising machine architecture for LDPC decoding
without adding auxiliary spins. 

\subsection {Co-designed formulation and architecture}\label{secc:m_qubo} 

To see how an Ising machine architecture can be extended to support LDPC
decoding, we rewrite the decoding objective function (Eq.~\ref{eq::obj})
using spin representation to more easily map to the hardware implementation,
specifically $\Tilde{C_i}=0\rightarrow\sigma_i=1$, and 
$\tilde{C_i}=1\rightarrow \sigma_i=-1$.
With this rewrite, an XOR operation on a binary variable is converted into 
a multiplication operation of the corresponding spin variable as can be
seen in Table~\ref{tabel:truth_table}. 
Thus Eq.~\ref{eq::obj} can be written as follows:
\begin{equation}
\label{eq::mul_obj}
\begin{split}
    f = \sum_{i=0}^{n-1}(R_i - \sigma_i)^2  + \alpha \sum_{j=0}^{m-1}(1-\prod_{\mathclap{ H_{j,i}=1}} \sigma_i)/2.
\end{split}
\end{equation}
Since constants do not affect the objective function, we can remove them:
\begin{equation}
\label{eq::LDPC_ising}
\begin{split}
    f =  \sum_{i=0}^{n-1}-2R_i\sigma_i  -\frac{\alpha}{2}\sum_{j=0}^{m-1} 
    \sigma_i \sigma_{j\backslash i},\;\;
    where\ \sigma_{j\backslash i}\ =\ \prod_{\mathclap{H_{j,k}=1, k\neq i} } \sigma_k
\end{split}
%\sum_{\mathclap{\;\;\;H_{j,i}=1}} 
\end{equation}
%\todo{check the above formulation}

\begin{wraptable}{r}{1.5in}
%\vskip -20pt
\footnotesize
\centering
\caption{Truth table of bit XOR operation and spin multiplication  }
\setlength{\tabcolsep}{2pt}
\label{tabel:truth_table}
\begin{tabular}{|c|c|c|c|c|c|} 
\hline
$A$ &  $B$ &  $A \xor B$ & $\sigma_A$ & $\sigma_B$ & $\sigma_A.\sigma_B$ \\ 
\hline\hline
 0 &  0  & 0 & 1 & 1 & 1\\  
\hline
0  &  1 & 1 & 1 & -1 & -1\\   
\hline
1    & 0 & 1 & -1& 1& -1\\ 
\hline
1    & 1 & 0  & -1 & -1 & 1\\ 
\hline
\end{tabular}
\end{wraptable}

We can see that Eq.~\ref{eq::LDPC_ising} bears a significant resemblance to the Ising model 
(Eq.~\ref{eqn:Ising_w_field}) if we treat all $\sigma_{j\backslash i}$ as some (auxiliary) spins. 
Note that these spins are simply functions of regular spins.
We can thus design an \emph{augmented} Ising machine that uses extra logic to generate these
auxiliary spins which are coupled with regular spins. 

On the surface, the use of auxiliary spins would suffer the same drawbacks of using extra
spins discussed earlier (hardware cost and bloated state space). In reality, these spins  
are fundamentally different from the 
auxiliary spins resulted from Eq.~\ref{eq:unary_binary} in two important ways as follows
(quantitative results will be shown later in Sec.~\ref{sec:evaluations}).
\begin{enumerate}
    \item \textbf{Degree of freedom:} The auxiliary spins ($\sigma_{j\backslash i}$) do not have
    any additional degree of freedom. This means that fundamentally they do not increase the size
    of the state space. Consequently, they do not add to the difficulty of navigating the energy landscape.
    
    \item \textbf{Coupling complexity:} Another common drawback for using auxiliary spins is that the
    circuit complexity is dominated by couplers which scales quadratically with the number of spins. 
    Indeed, we can see in Eq.~\ref{eq::LDPC_ising} that there are many auxiliary spins needed. 
    Take only one checksum $j$ for instance and assume
    (for simplicity of argument) that it involves 10 regular spins $\sigma_{i1}$ to $\sigma_{i10}$. 
    Then each such spin requires its own auxiliary spins $\sigma_{j\backslash i1}$ to 
    $\sigma_{j\backslash i10}$. This repeats for every one of the $m$ checksums, easily leading
    to tens of thousands of spins for kilo-bit codes in theory. But in reality, as we will 
    discuss in detail later,
    there is significant regularity, structure, and sparsity in the coupling needed for
    real-world codes. As a result, a custom-designed Ising machine has very limited coupling
    complexity.
\end{enumerate}

With this new formulation,
we now discuss the design of a complete system (illustrated in Fig.~\ref{fig:arch}), 
staring with \ding{172} the overall architecture centering on the coupling array 
architecture (Sec.~\ref{ssec:arch}), followed by \ding{173} circuit designs for spins, bias units, 
parity units, and coupling units (Sec.~\ref{ssec:circuit}).

\subsection{Architecture of the augmented Ising machine}
\label{ssec:arch}

In principle, any Ising machine can be adapted according to Eq.~\ref{eq::LDPC_ising}. In practice,
because the auxiliary spins are logical function of regular spins, we use BRIM~\cite{afoakwa.hpca21}
as the baseline. Its CMOS-compatible, voltage-based spins are more convenient for our purpose
than qubit- or phase-based machines~\cite{dwave-timing, wang2019oim}.

We first discuss the coupling array as it is by far the dominant component, generally containing $O(n^2)$ couplers
for $n$ spins. By default, there is a dense array of $n\times n$ \emph{programmable}
couplers to map the coupling coefficients $J_{ij}$ in Eq.~\ref{eqn:Ising_w_field}. A naive mapping
of Eq.~\ref{eq::LDPC_ising} for a $(k,n)$ parity code will require $n$ regular spins for the code
and up to $mn$ auxiliary spins for the parity checks. Fortunately, 
a number of customizations can be done to greatly simplify the overall architecture. 

\subsubsection{\bf Restricted and regular coupling}

To see the opportunity, let us use a concrete example and
assume a 6-spin, 2-checksum ($j=0,1$) system  
with the following quadratic coupling: \begin{equation}
\color{lightgray}\sum_{j=0}^{m-1}(1-\prod_{\mathclap{ H_{j,i}=1}} \sigma_i) \rightarrow \color{black}
\underbrace{(1-\sigma_1\sigma_3\sigma_5)}_{\text{checksum } j=0}+\underbrace{(1-\sigma_2\sigma_4\sigma_6)}_{j=1}
\label{eq:chksum_ex}
\end{equation}
Nominally we need 6 auxiliary spins ($\sigma_{0\backslash 1}$,
$\sigma_{0\backslash 3}$,
$\sigma_{0\backslash 5}$,
$\sigma_{1\backslash 2}$,
$\sigma_{1\backslash 4}$,
$\sigma_{1\backslash 6}$), making a total of 12 spins (and about 144 couplers).
But the coupling is actually restricted. Only an auxiliary spin is coupled to
a regular spin, and only in one direction -- the output of the auxiliary spin determines the current inflow to a regular
spin and can thus potentially change it.

Furthermore, within one checksum, the multiple auxiliary spins (\eg $\sigma_{0\backslash 1}$,
$\sigma_{0\backslash 3}$, $\sigma_{0\backslash 5}$) are ultimately reflection of the 
same parity checksum (the XOR of bits 1, 3, and 5). 
Another important regularity is that the coupling strengths are all the same. Combined
together, this means that a single auxiliary spin is needed per checksum and each 
coupler can be now implemented via a simplified control circuit, which we discuss
in Sec.~\ref{ssec:circuit}). Finally, the coupler is only needed when a spin is part
of the checksum.

Combining these factors, in our running example, we end up requiring only 2 auxiliary spins
and 6 total couplers. In fact, there are additional opportunities because  most
standards use proto-graph-based codes which have additional structures 
(see Box 1) that we can exploit as follows.

\begin{figure}[t]
\begin{boxedminipage}{\columnwidth}
\textbf{\small Box 1. LDPC matrix structure for 5G }
\vskip 8pt
\parskip 0.8ex
%\scriptsize
%\footnotesize
\small
In this paper, we discuss LDPC for the latest standard called 5G new radio (NR)~\cite{tr20175g} by the 3rd-generation partnership project (3GPP) for wireless enhanced-mobile broadband (eMBB) communication. 

The 5G standard uses two base-graph (BG) matrices: BG1 and BG2, which are used to construct a parity check matrix. Fig.~\ref{fig:5g-nr} shows the structure of BG1. The blue dots in Fig.~\ref{fig:5g-nr} show nonzero elements of the matrix. It has 46 rows and 68 columns (BG2 matrix is 42$\times$52). Rows in the base graphs describe parity check equations. By default, BG1 generates an LDPC code with a rate equal to $1/3$, and BG2 generates an LDPC code with a rate equal to $1/5$. 

\begin{wrapfigure}{r}{1.8in}
    \vskip -5pt
    \hskip -5pt
    {\includegraphics[width=1.9in]{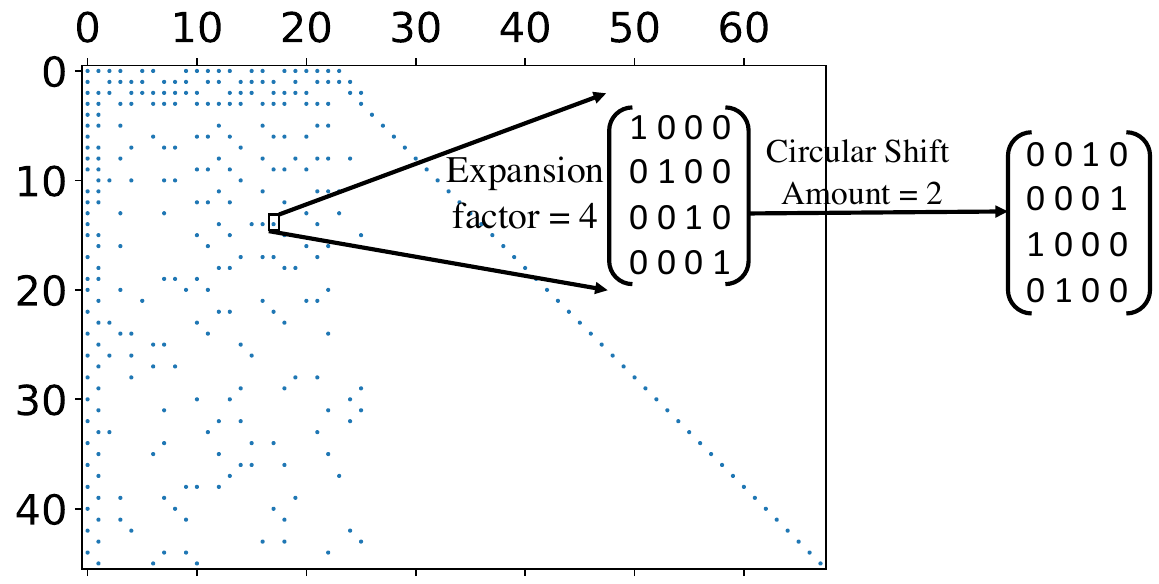}}
    \caption{Structure of 5G-NR standard base graph 1. Each blue dot represents a nonzero element which will be replaced by a circular identity matrix based on the expansion factor as shown.}
    \label{fig:5g-nr}
\end{wrapfigure}

Once the base graph is chosen, each zero entry is replaced with a $Z \times Z$ all-zero matrix. Each nonzero entry is replaced with a $Z \times Z $ circularly shifted identity matrix. $Z$ is called an expansion factor. The amount of circular shift depends on the value of the nonzero element. The values of the nonzero elements range from 0 to $Z-1$. If the value is 0, replace the element with the $Z \times Z$ identity matrix; otherwise, circularly shift the identity matrix based on the value. Fig.~\ref{fig:5g-nr} shows an example with an expansion factor of 4 and a circular shift amount of 2. The 5G-NR standard allows different discrete expansion factors ranging from 2 to 384. For further details, please refer to \cite{bae2019overview}.
\end{boxedminipage}
\end{figure}

\begin{figure}[htp]
    \centerline{\includegraphics[width=0.50\textwidth]{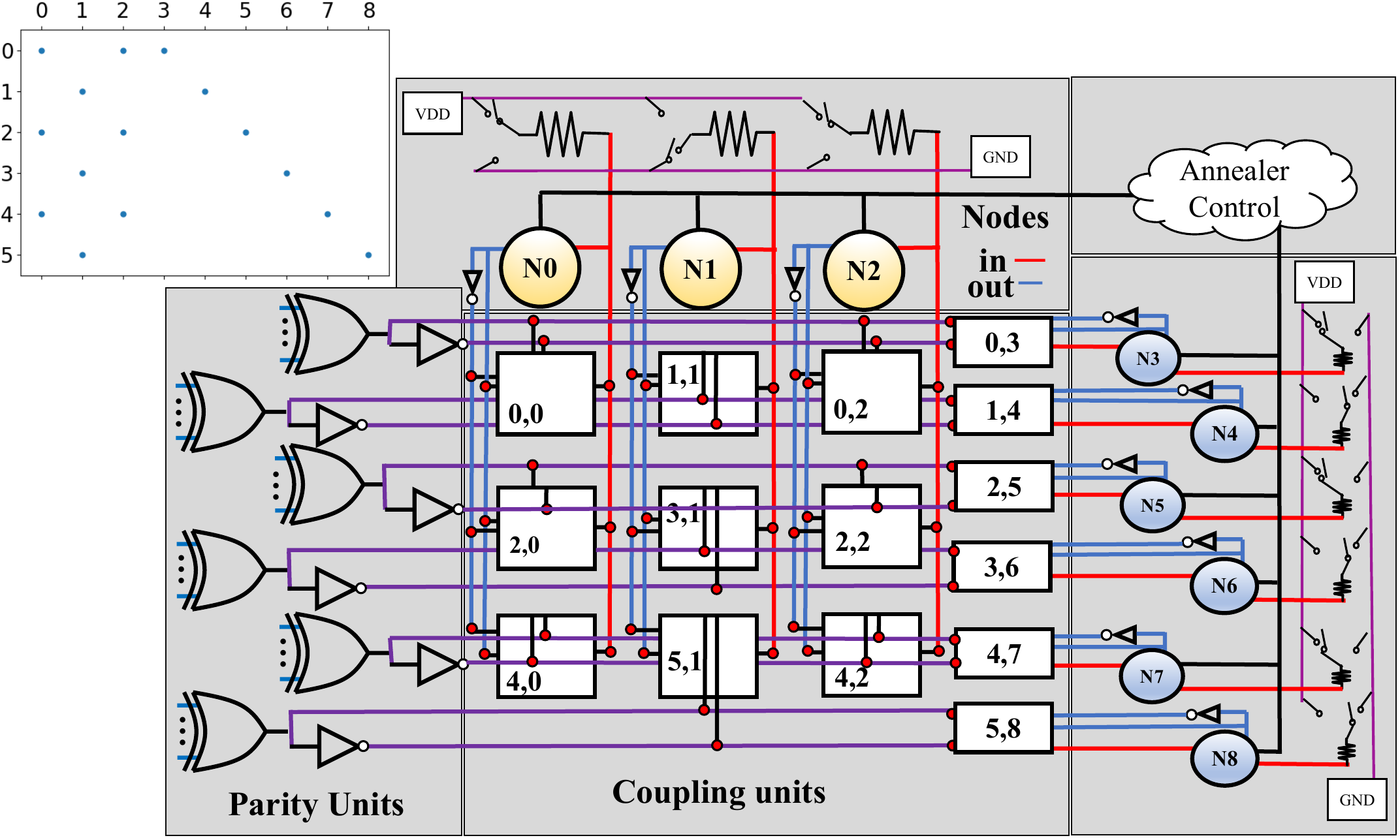}}
    \caption{Illustration of our Ising machine-based LDPC decoder using a specific portion 
    (shown in top left) of the parity check matrix.} \label{fig:arch}
    %\todo{need to add pseudo matrix }
\end{figure}

\subsubsection{\bf Structure of the matrix}
As the name low-density suggests, the matrix is rather sparse, reducing the number of couplers
needed. A particular property of the 5G base matrix is that a large portion of the code bits
are only involved in a single parity check (so-called extension checks). In other words, many
columns (representing bits) have only one non-zero element. This portion is organized as a
diagonal submatrix and can be seen in Fig.~\ref{fig:5g-nr}. Because these spins have only one
coupler, their couplers can be easily incorporated into their node design. Overall, the coupling
array can be organized as a 2-d grid with auxiliary spins (parity checks) as one dimension
regular spins as another, with the checksum bit having single parities forming a special column.
The structure is shown in Fig.~\ref{fig:arch}.

To recap, with this architecture, the number of couplers equals the number of 1s in the
parity check matrix and is thus a constant (given a matrix), not affected by the number of
auxiliary spins we added. The auxiliary spins are generated from the XOR of all bits involved
in a parity check and thus have no degrees of freedom on their own. The state space of the
dynamical system is only a function of the number of regular spins.

\subsection{Circuit design}
\label{ssec:circuit}

\subsubsection{\bf Nodes}

In our baseline Ising machine~\cite{afoakwa.hpca21}, the spins are represented 
by voltage polarities on capacitors. %which are terminated to a \emph{virtual} ground (at half $V_{dd}$).
When representing $\sigma_i=+1$, corresponding to bit $\tilde{C_i}=0$,
the capacitor has a positive voltage.
Conversely, for $\sigma_i=-1$, its voltage is negative.

\begin{wrapfigure}{r}{1.7in}
     {\includegraphics[width=1.7in]{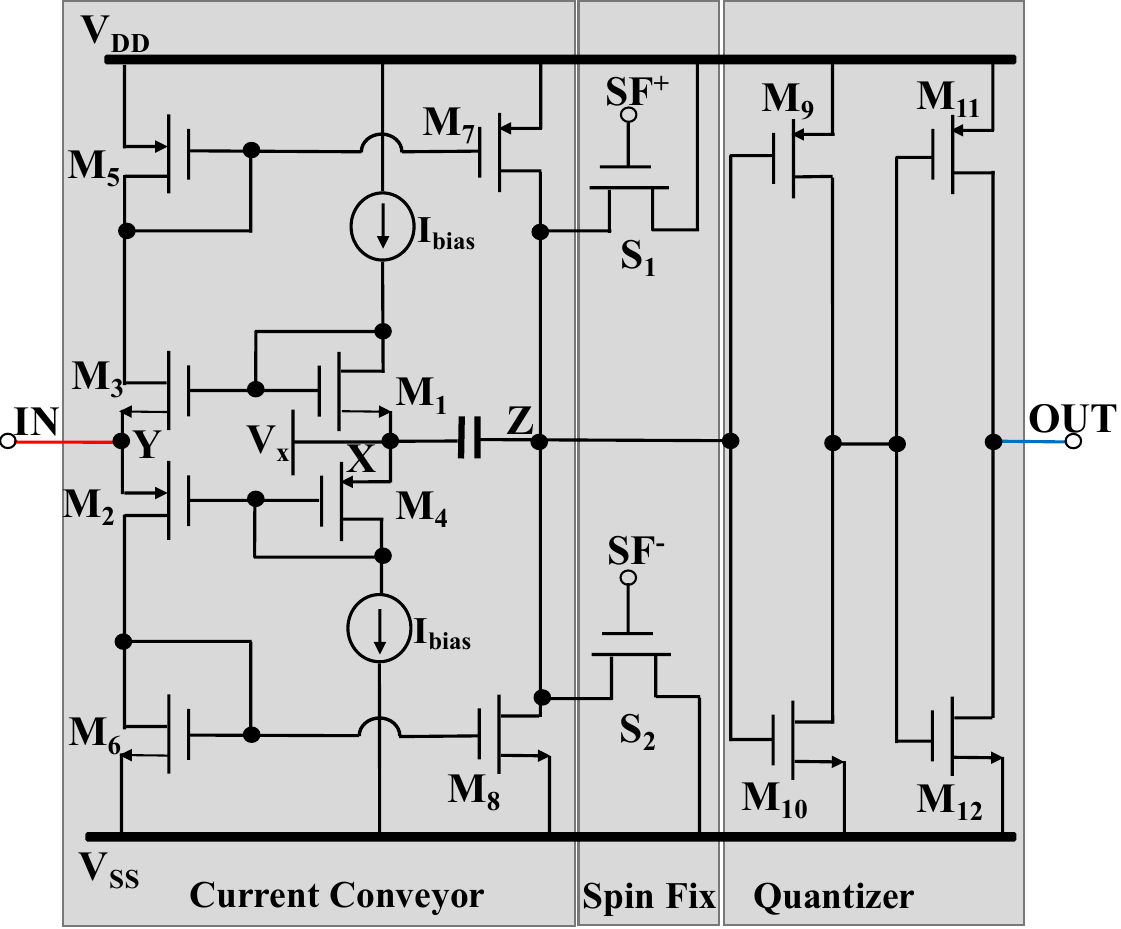}}
     \caption{Circuit implementation of Node of LDPC decoder.}
     \label{fig:QuBRIM_node}
\end{wrapfigure}

In the circuit implementation of the proposed Ising machine,
a spin is stored on a capacitor, and a node maintains the spin 
and interfaces with other circuits with its input/output circuit.
Fig.~\ref{fig:QuBRIM_node} shows an initial circuit design of a node. 
Besides the spin capacitor, it consists of three core functional blocks: 
a current conveyor, a one-bit quantizer, and a pair of spin-fix (SF) switches.
This node circuit takes an input current at $IN$, and generates an output voltage at $OUT$. 
The $OUT$ port of a node carries quantized voltage signals to both parity units and coupling units. 
The $IN$ port receives currents from coupling units. 
The node output can also be stored synchronously as a bit by an additional D flip-flop (DFF) when necessary.

%\begin{figure}[htp]
%\centering
%\includegraphics[width=0.30\textwidth]{LDPC codes/FIGS/qubrim_node.pdf}
%\caption{Circuit implementation of Node of LDPC decoder.} \label{fig:QuBRIM_node}
%\end{figure}

The current conveyor is constructed as a class-AB current mirror~\cite{kawahito1996cmos} %\cite{see Team QuiCC referenc}, 
which is designed for high slew rate and large input dynamic range. 
$M_5$-$M_8$ convey the input current from port $IN$ to node $Z$, 
hence charging/discharging the capacitor. 
The two current sources $I_{bias}$ set the bias condition for $M_1$-$M_4$, 
and hence the operating-point voltage at $Y$ is equal to that at $X$. 
By biasing node $X$ to $V_X=1/2(V_{DD}-V_{SS})$, 
$V_Y$ is effectively biased to the same voltage. 
The input impedance (at port $IN$) of this circuit is low,
allowing port $IN$ to operate as a current input.

The capacitor's voltage at node $Z$ represents the spin state: 
$V_Z<V_X$ for $\sigma_i=-1$, and $V_Z>V_X$ for $\sigma_i=+1$.
Note that this voltage $V_Z$ is an \emph{analog} signal, and needs to be quantized 
before it is used as an input for the XNOR gates in the parity and coupling units. 
Since our system operates in continuous-time, 
this quantizer can be simply constructed with inverters in series. 
In Fig.~\ref{fig:QuBRIM_node}, $M_{9-12}$ function as both a quantizer and super-buffer.

In order to configure the system's initial spins, 
%before the machine's states are allowed to evolve in seeking an energy minimum
two switches ($S_{1,2}$) are connected to node $Z$ to set/reset it to $V_{DD}$ or $V_{SS}$. 
We can also use these two switches to perform spin-fix perturbations, 
which allow the machine to escape local minima in search of the optimal solution. 
$SF_i^+$ and $SF_i^-$ are two non-overlapping control signals generated by the \emph{Annealer Control} unit.

\subsubsection{\bf Auxiliary spins and couplings}

For the auxiliary spins represent parity checks, 
let us consider the concrete case of checksum $j=0$ in Eq.~\ref{eq:chksum_ex}
($\tilde{C_1} \xor \tilde{C_3}\xor \tilde{C_5}$)
(the spin-based representation is $(1-\sigma_1\sigma_3\sigma_5)/2$). 
The physical meaning of the coupling is quite straightforward: 
when $\tilde{C_1}\xor \tilde{C_3}\xor \tilde{C_5}=0$,
each node will receive one unit of current keeping the node in the same polarity.
Conversely, when $\tilde{C_1}\xor \tilde{C_3}\xor \tilde{C_5}=1$, each node will
receive one unit of current trying to change the polarity of the node. 
We achieve this by generating the checksum ($\tilde{C_1}\xor\tilde{C_3}\xor \tilde{C_5}$)
to select for each spin whether to couple to its positive or negative edge of its own
node. An alternative view is that the selector acts as an additional XOR gate to
``back out" its own node. Thus, for node 1, the coupling comes from 
$\tilde{C_1} \xor \tilde{C_3}\xor \tilde{C_5}\xor\tilde{C_1}=\tilde{C_3}\xor \tilde{C_5}$.
%\begin{wrapfigure}{r}{3in}
%    {\includegraphics[width=3in]{LDPC codes/FIGS/circuit_diagram_ldpc_pass.pdf}}
%    \caption{Logical implementation of proposed LDPC decoder.}
%    \todo{show the new pass gate based coupling controlled by a single auxiliary spin. Also use the
%    same orientation as the matrix}
%    \label{fig:circuit_digaram}
%\end{warpfigure}

\begin{figure}[htp]
\centerline{\includegraphics[width=0.5\textwidth]{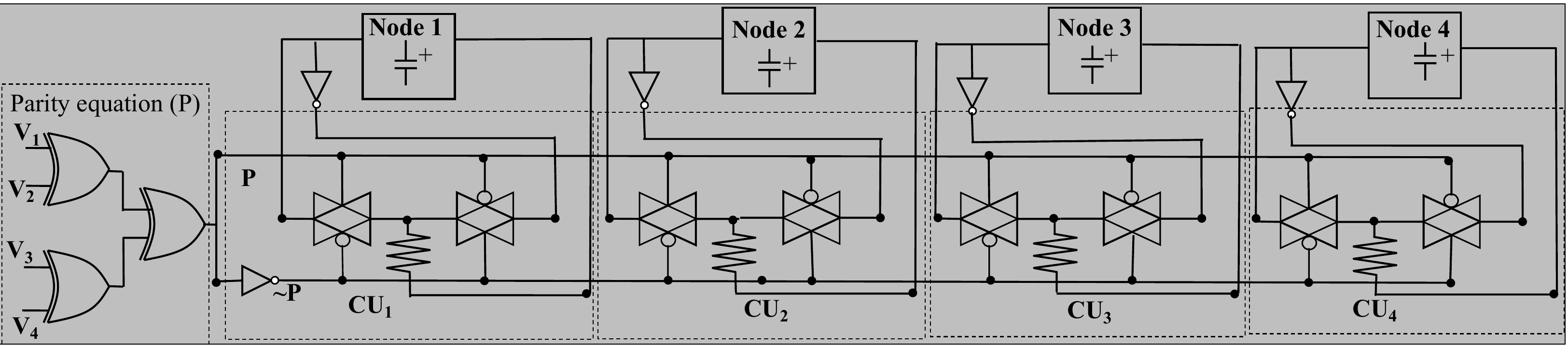}}
    \caption{Circuit implementation of parity check and its coupling to spins.} \label{fig:circuit_digaram}
\end{figure}

\subsubsection{\bf Bias}
The linear terms ($-2R_i\sigma_i$ in Eq.~\ref{eq::LDPC_ising}) helps to 
constrain the decoded output closer to the received message, and is implemented by a 
bias current at each node proportional to the received value $R_i$. 
More specifically, according to Eq.~\ref{eq::LDPC_ising}, the bias current for
node $i$ needs to be $\frac{4R_i}{\alpha}$ time that of the coupling current from
the parity check. The sign of $R_i$ determines the polarity of the bias current. 

Putting everything together, 
Eq.~\ref{eqn:rim_node_simplifed} describes the circuit behavior.
We can see that if we use voltage $v_i$ (relative to virtual ground) 
as an approximation for spin $\sigma_i$ in Eq.~\ref{eq::LDPC_ising}, 
Eq.~\ref{eqn:rim_node_simplifed} is proportional to the gradient.
\begin{equation}\label{eqn:rim_node_simplifed}
\begin{split}
%    \dv{v_i}{t} = \frac{1}{C}(I_{in}) =  \frac{1}{C} \left[ \sum_{\mathclap{j}} J_{ij}*(P_{j\backslash i}) + {J_b}_i*({v_b}_i) \right]
    \dv{v_i}{t} = \frac{1}{C}(I_{in}) =  \frac{1}{C} \left[ \sum_{\mathclap{j}} J_{ij}*(P_{j\backslash i}) + {J_b}_i*({v_b}_i) \right] \\
    \dv{v_i}{t} = \frac{J}{C}(I_{in}) =  \frac{J}{C} \left[ \frac{4R_i}{\alpha}*{v_b}_i  + 
    \sum_{\mathclap{H_{j,i}=1}} \sigma_{j\backslash i} \right]
\end{split}
\end{equation}

%% file: Sections/4evaluation.tex
\section{Experimental Analysis}
\label{sec:evaluations}

In this section, we discuss the experimental analysis of our design by describing the experimental methodology and comparing our system to existing LDPC decoders in solution quality, throughput, and energy consumption. 
These decoders can be implemented using GPUs, FPGAs, or as ASICs and are often 
explicitly designed to a particular code. We chose 3 implementations representating 
the state of the art ASIC decoders for 5G standards to compare against.

Behavioral simulation of our LDPC decoder's is done by solving Eq.~\ref{eqn:rim_node_simplifed}, 
using MATLAB's nonstiff, single step, 5th-order differential solver (ode45). Circuit parameters are 
obtained using Cadence with 45nm Generic Process Design Kit (GPDK045). 
We also ran a state-of-the-art variant~\cite{Isakov}  of 
Simulated Annealing to obtain solution quality for LDPC decoding using the traditional
QUBO formulation.

\subsection{Comparison with existing LDPC decoders}
\begin{comment}
\begin{table*}[htb]
\centering
\caption{comparison of hybrid LDPC decoder with existing decoders}
\label{table:comparision}
%\resizebox{\columnwidth}{!}
{
%\setlength{\tabcolsep}{4pt}
\begin{tabular}{|c|c|c|c|c|c|} 
\hline
\bf {Design\textbackslash Decoder} & \thead{~\cite{schlafer2013new}}  & \thead{~\cite{lin202133}}   & \thead {~\cite{lee2022multi}}  & \thead{~\cite{thi2021low}} & \thead{this work}\\  
\hline\hline

CMOS Technology  &   65nm LVT  & 28nm   & 65nm  & 65nm & 45nm\\ 
\hline
Standard        &  IEEE802.11ad   & 5G-NR & 5G-NR & 5G-NR & 5G-NR \\ 
\hline
Decoding Algorithm & Min-Sum & \thead{ Normalized\\Min-Sum}   &  \thead{ offset\\Min-Sum} & \thead{Combined \\ Min-Sum} & \thead{hybrid \\ Ising model}  \\
\hline
Block size         &  672  &  \thead{all cases of 5G}  & \thead{all cases of 5G} & 3808 & 1088 \\
\hline
\thead{Decoding Latency per iteration\\ (nsec)} &  4  & 787 & 1530  & 125.33  & 220  \\
\hline
\thead{Max Throughput per iteration \\(GBs/sec)} & 160.8 & 33.2  & 17.06 & 30.38 & 4 \\
\hline
Max Power (mW)    & 5360 & 232  & 413 & 259 & 6.59 \\
\hline
Area ($mm^2$) &   12.09  & 1.97  & 5.74 & 1.49 & - \\
\hline
Energy (pJ/bit) per iteration  & 36.1 & 29  & 24.2 & 8.52 & 1.33\\
\hline
\thead{BER @ SNR 3 \\ exp\_factor = 16 \\ bit width =8bits }  &  5.69E-02 & \red{3.67E-02}  & 3.67E-02 & \red{3.67E-02} & 1.61E-02\\
\hline
\end{tabular}
}
\end{table*}

\todo{block size must be the same, increase either area or latency to compensate for the block size, equation}

\end{comment}

\begin{table}[htb]
\centering
\caption{Comparison of the proposed LDPC decoder with existing decoders, all parameters are adjusted to implement highest expansion factor}

%\donotshow{
%}

\label{table:comparision}
\resizebox{\columnwidth}{!}{
\setlength{\tabcolsep}{4pt}
\begin{tabular}{|c|c|c|c|c|} 
\hline
\bf {Design\textbackslash Decoder}  & \thead{~\cite{lin202133}}   & \thead {~\cite{lee2022multi}}  & \thead{~\cite{thi2021low}} & \thead{this work}\\  
\hline\hline

CMOS Technology    & 28nm   & 65nm  & 65nm & 45nm\\ 
\hline
Standard          & 5G-NR & 5G-NR & 5G-NR & 5G-NR \\ 
\hline
Decoding approach & \thead{ Normalized\\Min-Sum}   &  \thead{ offset\\Min-Sum \\ (OMS)} & \thead{Combined \\ Min-Sum} & \thead{Augmented \\ Ising machine}  \\
\hline
Block size         &  26112 & 26112 & 3808 (26112$^*)$  & 26112 \\
\hline
\thead{Decoding Latency per iteration\\ (nsec)}  & 787 & 1530  & 125.33 (860$^*)$  & 320  \\
\hline
\thead{Max Throughput per iteration \\(Gbs/sec)}  & 33.2  & 17.06 & 30.38 & 81.6 \\
\hline
Max Power (mW)     & 232  & 413 & 259 & 158.24 \\
\hline
Area ($mm^2$)  & 1.97  & 5.74 &  1.49 & 5.44 \\
\hline
Energy (pJ/bit) per iteration   & 29  & 24.2 & 8.52 & 1.94\\
\hline
quantization   & 5  & 8  & 4 & 8\\
\hline
%\thead{BER \\  SNR = 2 \\exp\_factor = 64}   & - & - & - &-\\
%\hline
\end{tabular}
}
\footnotesize *Latency scaled for direct comparison with other designs.
%\todo{change xxx and yyy to original paper numbers}
\end{table}

\subsubsection {Comparison with ASIC LDPC decoders}
Table~\ref{table:comparision} shows a detailed comparison of our work with existing ASIC LDPC decoders. 
The decoders are designed for 5G standards.~\cite{lin202133},~\cite{lee2022multi} support all the LDPC codes described by the 5G standards, whereas~\cite{thi2021low} is designed for a particular code created by a specific expansion factor.~\cite{thi2021low} reported their parameters for a decoder intended for an expansion factor of 56. We have modified its decoding latency to represent the decoder designed for an expansion factor of 384 by keeping area and power constant—all the other parameters are directly obtained from the corresponding papers. The table shows that our physical computation-based decoder consumes 4.4 times less energy per bit than the state of the art. Finally, earlier work using Ising machine for LDPC decoding achieves only 21Mb/s~\cite{kasi2020towards}.

\subsection{Comparison with traditional QUBO Formulation }\label{sec:solution_quality}

One thousand messages were encoded using different LDPC codes to obtain the solution quality. These codes were constructed using the 5G-NR standard's base graph 1 with varying expansion factors. The encoded messages were then modulated using BPSK modulation. Noise was added to the modulated signals based on the signal-to-noise ratio (SNR). These corrupted signals were used to test the different algorithms for solution quality. 

\begin{figure}[htb]
 \begin{minipage}[b]{.6\columnwidth}
   \includegraphics[width =\textwidth]{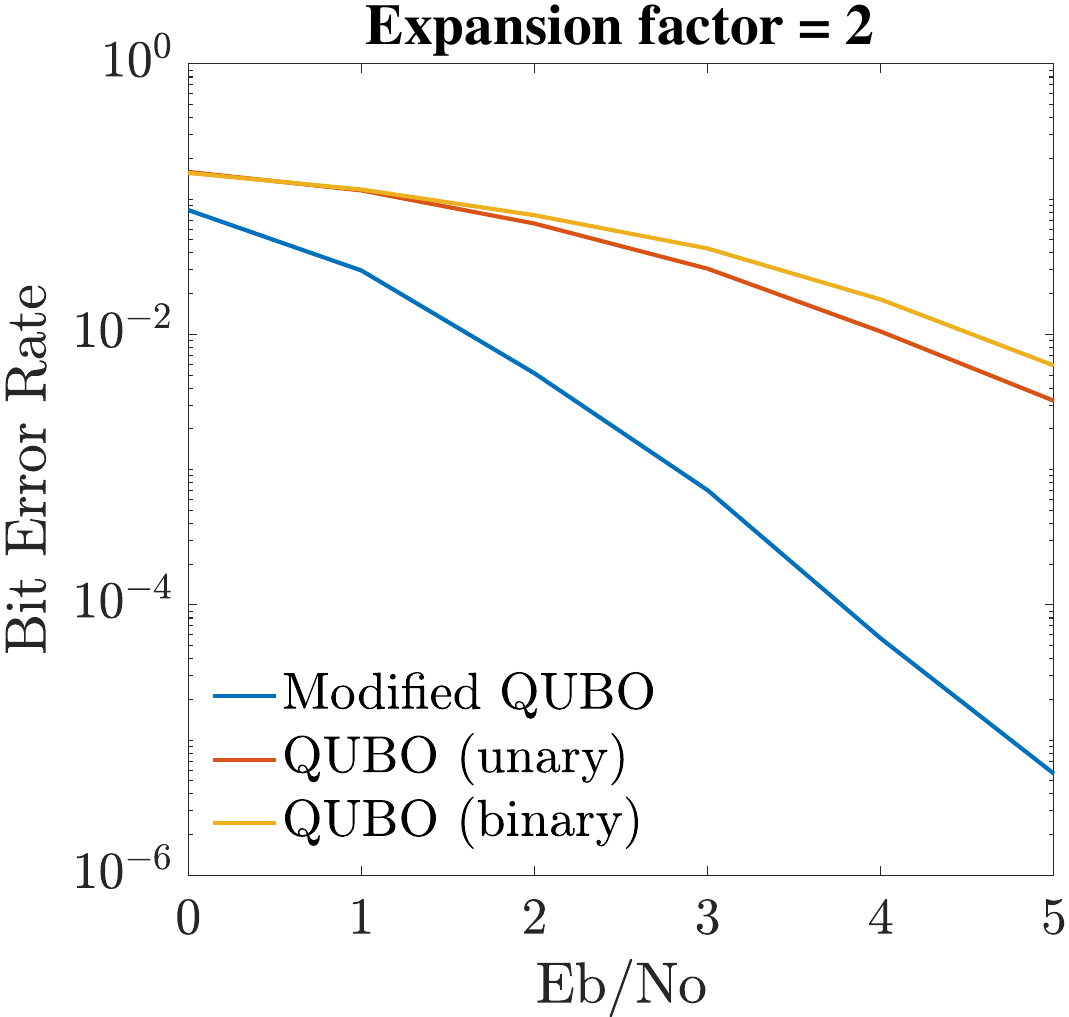}
  %\caption{This is first figure}
  \end{minipage}%
  \hfill
  \begin{minipage}[b]{.35\columnwidth}
  \includegraphics[width = \columnwidth]{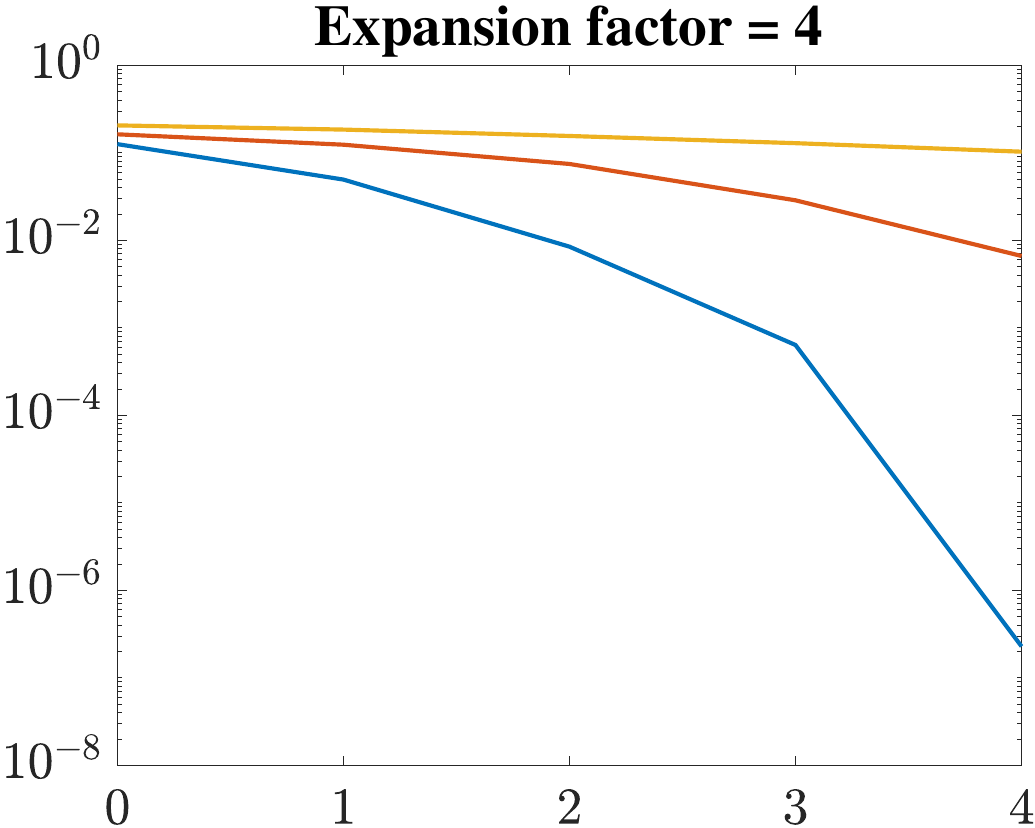} 
  \includegraphics[width = \columnwidth]{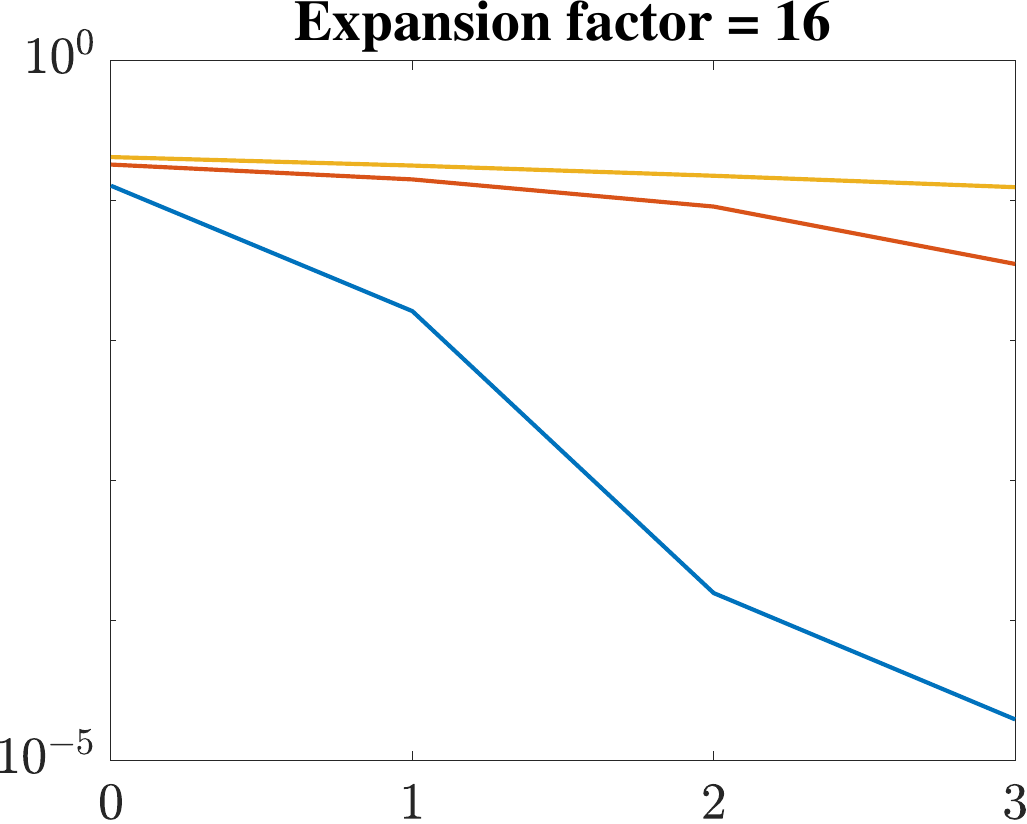}
  %\caption{caption}
  \end{minipage} 
  \caption{Bit Error Rate (BER) achieved by different algorithms at different Signal-noise ratios (Eb/No). Lower is better.}
\label{fig:ber}
\end{figure}

Fig.~\ref{fig:ber} shows the solution quality of LDPC decoding using standard QUBO formulations and our proposed formulation (hybrid Ising model or modified QUBO). Since we are analyzing the
quality of the formulation, we use the much faster 
Simulated Annealing (for 10000 iterations) to measure BER. Each message was annealed 10 times with random initializations. Fig.~\ref{fig:ber} shows the average of the bit error rate obtained. We can observe from the graphs that, given the runtime, both the formulations of QUBO yield worse BER compared to our proposed formulation. 

The better performance for our formulation is part of the reason our co-designed
Ising machine performs much better than earlier work using D-Wave annealer.
Additionally, our proposed architecture requires far fewer couplers:
20K couplers (expansion factor 64) vs 
443K and 290K couplers for unary or encoding requires  respectively.

%We find the distinct solutions in these different anneals and rank them according to their Ising energies. Considering this order static, the \textit{expected} Bit Error rate (E(BER($N_a$)))~\cite{kim2019leveraging} for each instance is calculated using the formula~\ref{eqn:expected_ber}.

%\begin{equation}\label{eqn:expected_ber}
%\begin{split}
%E(BER(N_a)) = \sum_{q=1}^{L} \left[ (\sum_{r=q}^{L} P(r))^{N_a} -  (\sum_{r=q+1}^{L} P(r))^{N_a}\right] \\ F(q)/K   
%\end{split}
%\end{equation}
%where K is the number of bits in the message, L is the number of unique solutions obtained by different anneals, r is the rank index of each solution, $P(r)$ is the probability of obtaining the $r_{th}$ solution, $F$ is the number of bit errors in the $q^{th}$ solution, relative to the ground truth and $N_a$ represents number of anneals. The average of 1000 different expected BER is reported as the final Bit Error rate for each SNR. As shown in the figure, modified QUBO gives us the better BER when compared to QUBO formulation.

\begin{comment}
\begin{figure}[htb]
 \begin{minipage}[b]{.6\columnwidth}
   \includegraphics[width =\textwidth]{FIGS/qubo_2.pdf}
  %\caption{This is first figure}
  \end{minipage}%
  \hfill
  \begin{minipage}[b]{.35\columnwidth}
  \includegraphics[width = \columnwidth, height=0.72\columnwidth]{FIGS/qubo_4.pdf} 
  \includegraphics[width = \columnwidth, height=0.72\columnwidth]{FIGS/qubo_16.pdf}
  %\caption{caption}
  \end{minipage} 
  \caption{Bit Error Rate (BER) achieved by different algorithms at different Signal-noise ratios (Eb/No). Lower is better.}
\label{fig:ber}
\end{figure}
\end{comment}

%\todo{needs to be changed}

\begin{figure}[htb]
  \centering
   \includegraphics[width=0.50\textwidth]{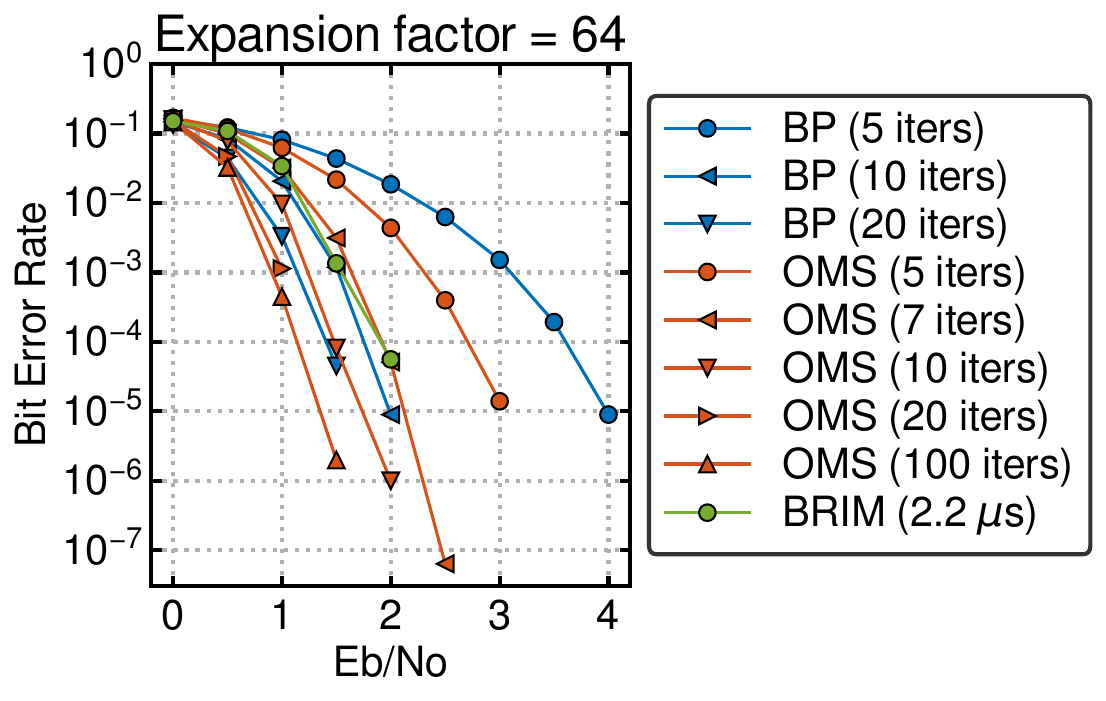}
  \caption{Bit error rate achieved by different algorithms at different Signal-to-noise ratios. Lower is better.}
\label{fig:ber_brim}
\end{figure}

To compare the solution quality of our proposed design with different variants of belief propagation algorithm, we encoded ten thousand messages using LDPC code constructed using the 5G-NR standard's base graph 1 with an expansion factor of 64. Fig.~\ref{fig:ber_brim} shows the bit error rate obtained by different algorithms. `BP' and `OMS' in the legend represent layered Belief Propagation and layered Offset Min-Sum algorithm, respectively. 
BP and offset Min-sum algorithms were implemented using MATLAB's 5G toolbox. 
We simulate our system evolving for 2.2 $\mu s$. The graph shows that the BER obtained by the hardware approaches the BER rate obtained by BP and its variants at the 7th iteration.

%% file: Sections/5conclusion.tex
\section{Conclusions}\label{sec:conclusion}
%\todo{needs to be changed}

Low density parity code (LDPC) has become the standard channel coding for many communication applications, including the 5G-NR. Fast and energy efficient LDPC decoding is thus crucial for  modern communication infrastructure and the society that depends on it. The vast majority of research and prototyping efforts in decoding LDPC are focused on engineering an efficient custom circuitry hard-wiring a variant of a Belief Propagation (BP) algorithm. An alternative approach is to express the decoding as a Combinatorial Optimization Problem (COP) and leverage recent hardware advances that accelerate COP solving. However, traditional QUBO formulations of LDPC
decoding require many auxiliary spins with their own degrees of freedom, thus
increasing the problem state space and negatively affecting the solution quality given a
fixed annealing time budget.
This paper shows a new way of solving LDPC decoding by co-designing the hardware and formulation.
The new approach significantly improves solution quality over traditional formulations.
Enabled by a state-of-the-art Ising machine architecture, our proposed hardware is estimated
to achieve a 4.4 times reduction in energy per bit compared to the most efficient
decoder in literature. 

%still need to investigate if the proposed design would result in a better LDPC decoder compared to the state of the art. %Our preliminary results show that the proposed modification helped to achieve an LDPC decoder with a throughput of 118 Gbps and power consumption of 158.24mW, representing a 6x reduction in energy per bit compared to state-of-the-art decoders for 5G standards.

\begin{comment}
\begin{itemize}
    \item  LDPC codes are important ECC codes for its ability to reach the Shannon limit
    \item they have been standardized for many different applications
    \item The decoding of LDPC codes is a combinatorial optimization problem
    \item Ising machines have shown to solve COPs very effectively
    \item major problem: to map cop with constraints on ising machines needs to be converted into a QUBO which requiers some extra spins
    \item this makes the state space for search increase exponentially
    \item with changes to the hardware of BRIM we can directly map decoding LDPC without conversion
    \item this leads to design of LDPC decoder with \blue{118.6 Gbits/sec} through, \blue{158.24mW} power consumption.
    \item this design helps in reduction of energy consumption per bit by an order of magnitude compared to other state of art LDPC decoders
    
\end{itemize}
\end{comment}